\documentclass[12pt]{article}

\usepackage{graphicx}
\usepackage{verbatim}   
\usepackage{color}      
\usepackage{subfigure}  
\usepackage[abs]{overpic}
\usepackage{float}
\restylefloat{table}
\usepackage{color}
\usepackage{array}
\usepackage{rotating}
\usepackage{subfloat}

\usepackage{longtable}

\usepackage{multirow}

\newcommand\pubnumber{ATL-PHYS-PROC-2021-016}
\newcommand\pubdate{\today}

\def\institute{
INFN Gruppo Collegato di Udine, Sezione di Trieste, Udine and ICTP, Trieste \\
Strada Costiera 11, Trieste 34151, Italy}

\def\Title#1{\begin{center} {\Large #1 } \end{center}}
\def\Author#1{\begin{center}{ \sc #1} \end{center}}
\def\Address#1{\begin{center}{ \it #1} \end{center}}

\newcommand\pubblock{\rightline{\begin{tabular}{l} \pubnumber\\
         \pubdate  \end{tabular}}}
\newenvironment{Abstract}{\begin{quotation}  }{\end{quotation}}
\newenvironment{Presented}{\begin{quotation} \begin{center} 
             PRESENTED AT\end{center}\bigskip 
      \begin{center}\begin{large}}{\end{large}\end{center} \end{quotation}}

\def\ttbar{\ensuremath{t\bar{t}}}
\def\pt{\ensuremath{p_{\mathrm{T}}}} 

\def\beq{\begin{equation}}
\def\eeq#1{\label{#1}\end{equation}}
\def\eeqn{\end{equation}}

\def\beqa{\begin{eqnarray}}
\def\eeqa#1{\label{#1}\end{eqnarray}}
\def\eeqan{\end{eqnarray}}

\let\bar=\overbar

\def\Dslash{\not{\hbox{\kern-4pt $D$}}}
\def\dslash{\not{\hbox{\kern-2pt $\del$}}}

\def\msb{{\bar{\ssstyle M \kern -1pt S}}}

\begin{document}
\begin{titlepage}
\pubblock

\vfill
\Title{Treatment of top-quark backgrounds in extreme phase spaces: \\ the ``top $\pt$ reweighting'' and novel data-driven estimations in ATLAS and CMS
}
\vfill
\Author{ \textbf{Leonid Serkin} \\ \normalsize{on behalf of the ATLAS and CMS Collaborations\footnote{Copyright [2021] CERN for the benefit of the ATLAS and CMS Collaborations. Reproduction of this article or parts of it is allowed as specified in the CC-BY-4.0 license}}}
\Address{\institute}
\vfill

\begin{Abstract}

The top quark plays an important role in searches for physics beyond the SM, both as a dominant background and as a key signature for the signal. The most notable feature found in the top physics analyses in both ATLAS and CMS Collaborations: the disagreement between simulation and data of the top quark $\pt$ spectrum - is highlighted. A reweighting procedure which significantly improves the agreement between simulation and data, also known as the ``top $\pt$ reweighting'', is summarised. Commonly raised points concerning the reweighting to fixed-order predictions are discussed, and several refined approaches are mentioned. An overview of several data-driven methods developed and used to estimate the $\ttbar$ background in regions with large jet and $b$-jet multiplicities and/or high top quark $\pt$ is presented. 
\end{Abstract}

\vfill
\begin{Presented}
$13^\mathrm{th}$ International Workshop on Top Quark Physics\\
IPPP Durham (online version), UK, September 14--18, 2020
\end{Presented}
\vfill
\end{titlepage}
\def\thefootnote{\fnsymbol{footnote}}
\setcounter{footnote}{0}

\normalsize 

\section{Introduction}
With a mass close to the scale of electroweak symmetry breaking, the top quark, besides having a large coupling to the Standard Model (SM) Higgs boson~\cite{bib:ttH_CMS,bib:ttH_ATLAS}, is predicted to have large couplings to new particles hypothesized in many models beyond the Standard Model (BSM). Possible new phenomena may enhance the cross sections over SM predictions for various processes involving or decaying into top quarks, as detailed in Ref.~\cite{bib:me} and references therein. Hence, the top quark plays an important role in searches for BSM physics, both as dominant background and as key signature for signal.

\medskip
\noindent
In the following, the reweighting procedure which significantly improves the agreement between simulation and data, also known as the ``top $\pt$ reweighting'', is discussed. Recent results by the ATLAS~\cite{bib:ATLAS_det} and CMS~\cite{bib:CMS_det} experiments at the LHC using $\sqrt{s}=13$ TeV data that apply the top $\pt$ reweighting are summarised. Several issues concerning the reweighting to fixed-order predictions are mentioned and future approaches are discussed. Secondly, novel data-driven techniques for the estimation of the $\ttbar$ production background are presented. The focus is made on analyses using $\sqrt{s}=13$ TeV data and exploring high $\pt$ and/or large multiplicity regimes.

\section{Top quark $\pt$ reweighting}
\begin{figure*}[h]
\centering
\begin{overpic}[height=0.3\textwidth]{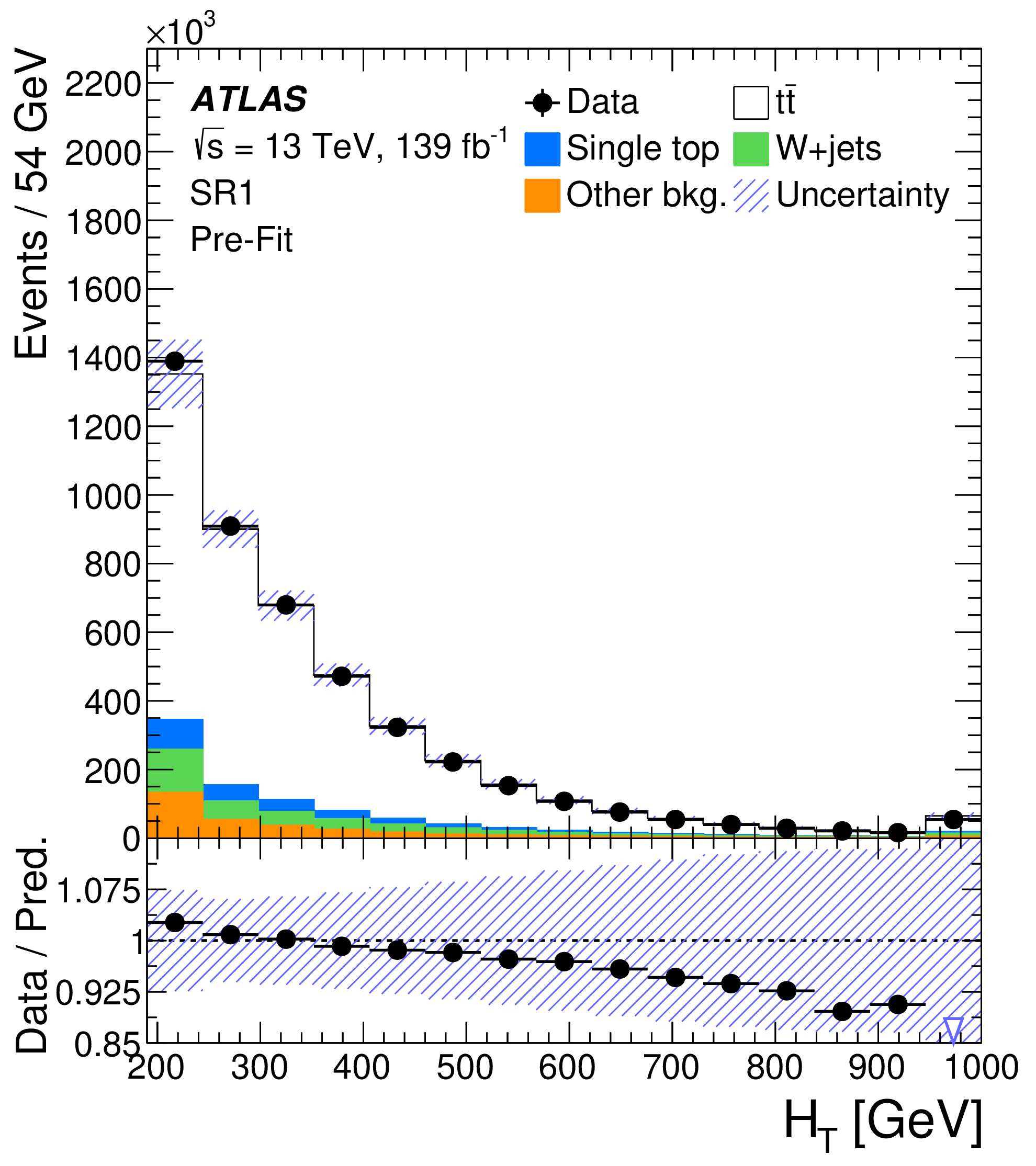}
\put(60,-7){(a)}
\end{overpic}
\qquad
\qquad
\begin{overpic}[height=0.3\textwidth]{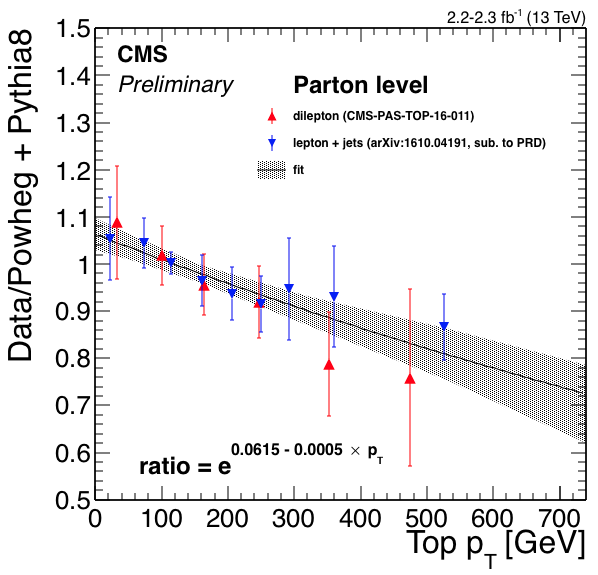}
\put(80,-7){(b)}
\end{overpic}
\caption{
(a) A clear slope due to the top quark $\pt$ mismodelling in the distribution of the scalar sum of jet $\pt$~\cite{bib:ttXsec_ATLAS}. (b) Ratio between 13 TeV data and Powheg simulation for the differential top quark pair cross section as a function of $\pt$ of the top quark~\cite{bib:cms_summary}. 
}
\label{fig:PlotN} 
\end{figure*}

One of the most notable features found in the top physics analyses is the disagreement between simulation and data of the top quark pair production. In particular, the $\pt$ distribution of the top quark is softer than predicted by MC event generators using NLO-accurate matrix elements with parton showering (denoted as NLO MC+PS), a trend observed in both resolved and boosted regimes and consistent among ATLAS and CMS . This mismodelling, first observed around 2012 in differential $\ttbar$ cross-section measurements (see Ref.~\cite{bib:TOP2013} and references therein), largely affects the data/MC agreement, as seen in Fig.~\ref{fig:PlotN}(a), and hence is one of the main uncertainties limiting precise measurements.

\medskip
\noindent
A short-term solution for this mismodelling was proposed in Refs.~\cite{bib:CMS_toprew,bib:ATLAS_toprew}: an empirical reweighting for top quark pairs based on the $\pt$ spectrum of generated-level top quarks, the so-called ``top $\pt$ reweighting''. Initially, the reweighting was applied to simulated $t\bar{t}$ events based on the ratio of measured differential cross sections~\cite{bib:ATLAS_diff,bib:CMS_diff1,bib:CMS_diff2} between data and the simulation as a function of top quark $\pt$, as seen in Fig.~\ref{fig:PlotN}(b). With the advent of the differential cross sections computed at next-to-NLO accuracy in pQCD (denoted as NNLO QCD)~\cite{bib:Czakon2}, and those including electroweak (EW) corrections computed at NLO accuracy (denoted as NNLO QCD + NLO EW)~\cite{bib:Czakon}, a reweighting of the $t\bar{t}$ samples in order to match their top quark $\pt$ distributions to that predicted by the best available fixed-order calculation became available.

\medskip
\noindent
Besides significantly improving the agreement between simulation and data, the top $\pt$ reweighting is also able to reduce the size of the modelling systematics affecting the simulation, improving the sensitivity as well as the reliability of the systematics model. The treatment of the top $\pt$ reweighting in both ATLAS and CMS analyses changes depending on the use case, i.e. correcting efficiencies or acceptance in $t\bar{t}$ measurements or searches. As an example, the reweighting of parton-level kinematics to differential cross-section measurements is not recommended for searches in high $\pt$ regimes due to the lack of statistics in deriving the weights in the tails. If no control region in data can be identified during a search, the analysis in CMS would correct the nominal sample to the best available fixed-order prediction by default, while ATLAS would consider it as systematic uncertainty. A summary of recent ATLAS and CMS searches and measurements applying different top $\pt$ reweightings and their associated uncertainties are presented in Table~\ref{tab:toppt}.

\begin{table}[t!]
\caption{A non-comprehensive list of recent ATLAS and CMS searches and measurements applying different top $\pt$ reweightings and their associated uncertainties. 
}
\begin{center}
\begin{tabular*}{\textwidth}{m{12em} m{13em} m{12em} } 
  \hline
  \hline
\multicolumn{1}{c}{Analysis / search }
 & \multicolumn{1}{c}{Top $\pt$ reweighting} &  \multicolumn{1}{c}{Uncertainty}   \\ 
  \hline
SM $t\bar{t} t\bar{t}$ \newline 1 and 2 OS leptons \newline CMS Collaboration~\cite{bib:4tops_CMS}    & Function derived from differential measurements & Allowing the correction function to vary within a $\pm 1\sigma $ uncertainty \\
  \hline
BSM $t\bar{t} t\bar{t}$ \newline 0 and 1 lepton final state \newline ATLAS Collaboration~\cite{bib:VLQ_ATLAS}    &   NNLO QCD  & Full difference between applying and not applying
 \\
\hline
SM $t\bar{t}H$ \newline multi-lepton final state \newline CMS Collaboration~\cite{bib:ttH_ML_CMS} 
&   NNLO QCD + NLO EW   & Difference between nominal and reweighted sample \\
 \hline
$t\bar{t}$ cross-section \newline 1 lepton final state \newline ATLAS Collaboration~\cite{bib:ttXsec_ATLAS}  &   NNLO QCD + NLO EW  & Symmetrised full difference between nominal and the reweighted sample \\
  \hline
  \hline
\end{tabular*}
\label{tab:toppt}
\end{center}
\end{table}

\begin{figure*}[t!]
\centering
\begin{overpic}[height=0.3\textwidth]{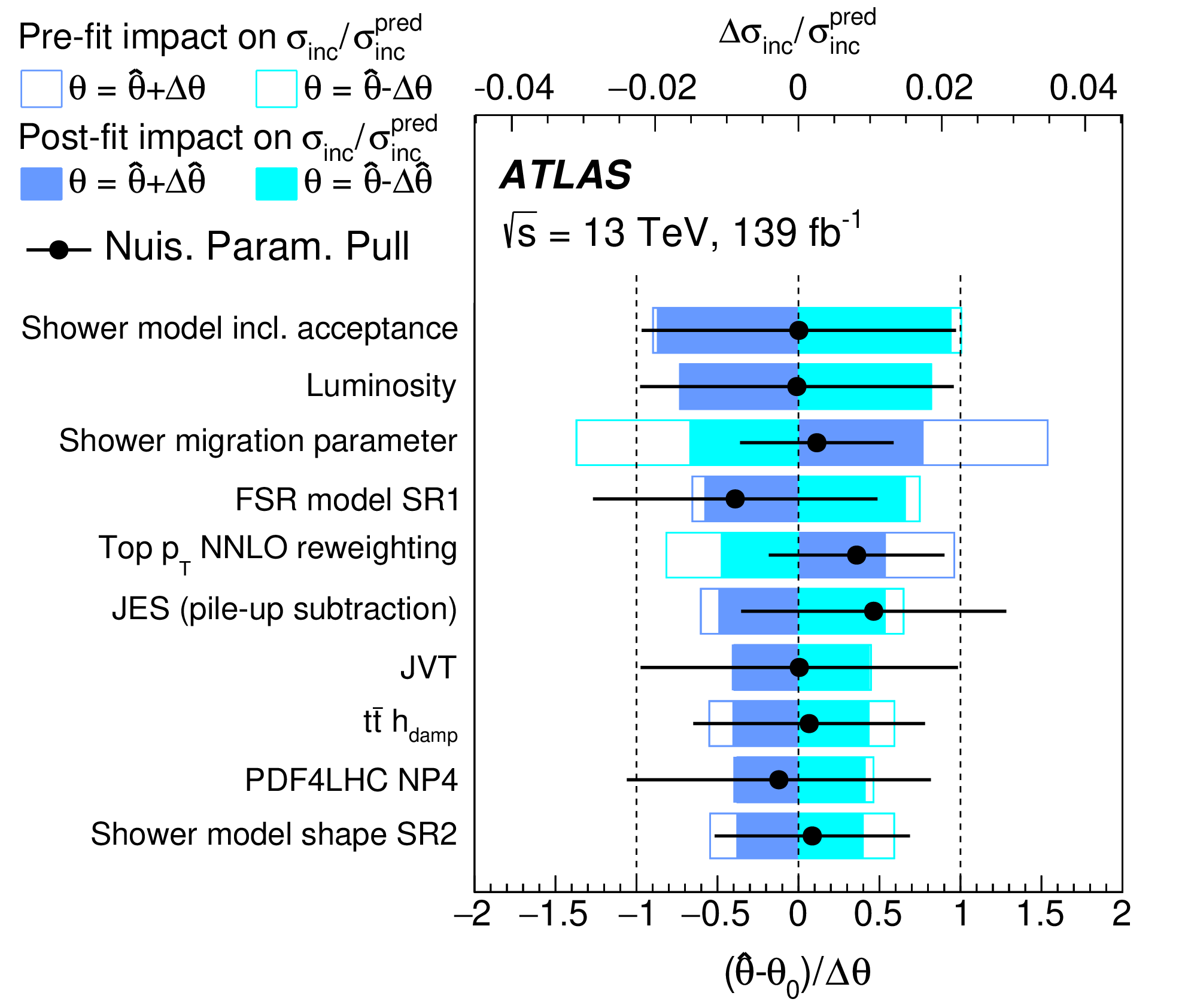}
\put(90,-9){(a)}
\end{overpic}
\qquad
\begin{overpic}[height=0.3\textwidth]{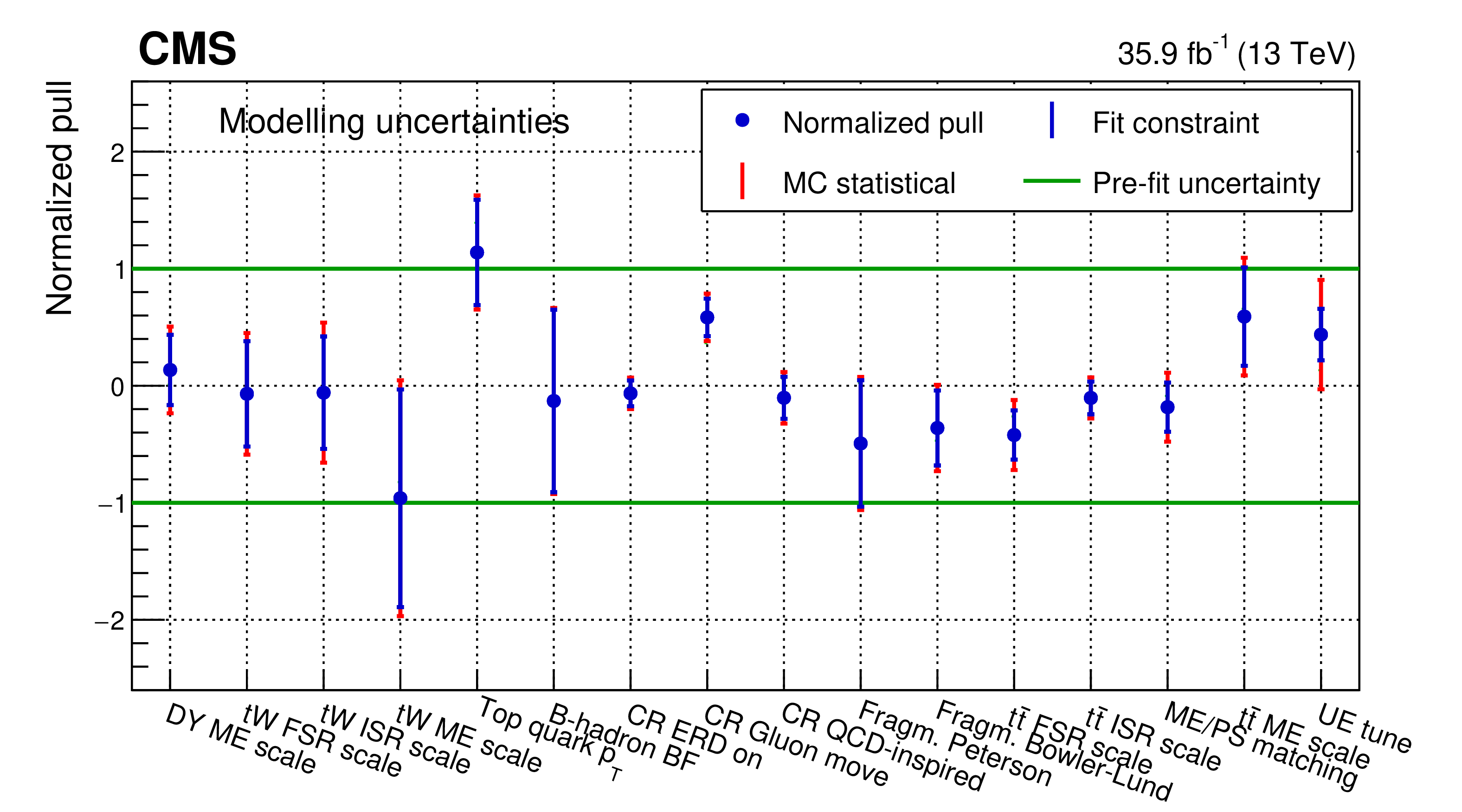}
\put(110,-9){(b)}
\end{overpic}
\caption{
(a) Ranking plot showing the effect of the ten most important systematic uncertainties in the 13 TeV $t\bar{t}$ cross-section measurement by ATLAS~\cite{bib:ttXsec_ATLAS}. The top $\pt$ reweighting parameter (5th from the top) is pulled towards the fixed-order prediction, as expected.
(b) Normalised pulls and constraints of the nuisance parameters for the 13 TeV $t\bar{t}$ cross-section measurement by CMS~\cite{bib:ttbarXS_CMS}. The nuisance parameter for the $\pt$ distribution of the top quarks (5th from the left) is pulled by one standard deviation, as expected.
}
\label{fig:Plot1} 
\end{figure*}

\medskip
\noindent
Usually both ATLAS and CMS assign a systematic uncertainty derived from the difference between the applying and not the top $\pt$ reweighing; if used in a maximum likelihood fit, this nuisance is profiled and is commonly accepted that the nuisance parameter responsible for this corrections could be pulled indicating the need for this correction in data (see Fig.~\ref{fig:Plot1}).

\subsection{Considerations on reweighting to fixed-order predictions}
\begin{figure*}[h]
\centering
\begin{overpic}[height=0.3\textwidth]{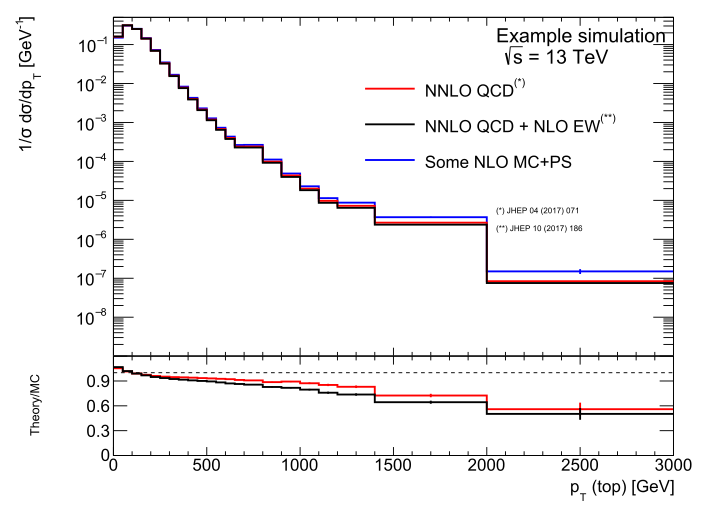}
\put(90,-7){(a)}
\end{overpic}
\qquad
\begin{overpic}[height=0.3\textwidth]{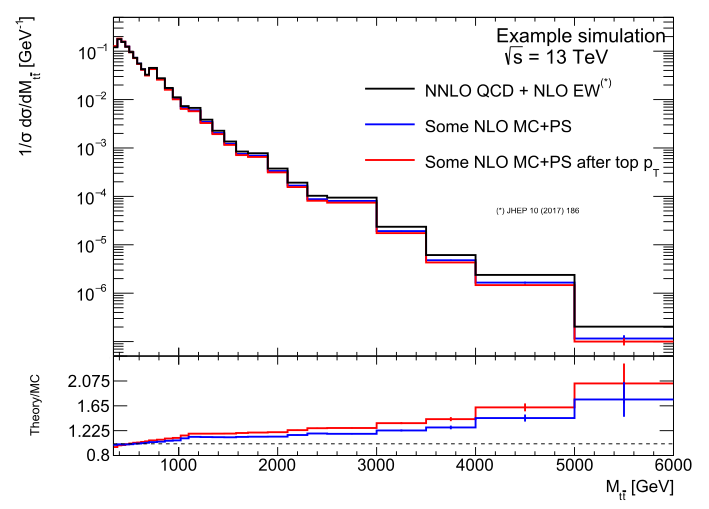}
\put(90,-7){(b)}
\end{overpic}
\caption{
(a) Distribution of the top quark $\pt$. The bottom panel shows the ratio of an example NLO MC+PS simulation with respect to the NNLO QCD prediction without (red) and with (black) NLO EW corrections. 
(b) Distribution of the $t\bar{t}$-system mass. The bottom panel shows the ratio of the NNLO QCD + NLO EW prediction with respect to an example NLO MC+PS simulation with (red) and without (blue) applying the top $\pt$ reweighting, derived as the black line in Fig.~\ref{fig:Plot2}(a). 
Both distributions are normalised to unity.
}
\label{fig:Plot2} 
\end{figure*}

The commonly raised points concerning the reweighting to fixed-order predictions are summarised in the following:
\begin{itemize}
\item The latest fixed-order calculations recommend different functional forms for the renormalisation and factorisation scales for different observables; the parton distribution functions (PDF), the top mass settings as well as the scale variations and scale choices are not always easily available in the fixed-order predictions, nor match the ATLAS and CMS settings for the MC generation;
\item In the current way the top $\pt$ reweighting is applied, only the top $\pt$ differential distribution is corrected to the given fixed-order calculation: yet, the question about other (partially correlated) variables (e.g. $t\bar{t}$-system mass) remains open (see Fig.~\ref{fig:Plot2}). After the top $\pt$ reweighting is applied, analysers within both Collaborations are requested to monitor the change in agreement between data and the corrected MC in other distributions and make sure that the reweighting does not spoil the initial agreement. 

\end{itemize}

\medskip
\noindent
Several refined approaches to the reweighting are possible, and a list of options to be explored is presented below:

\begin{itemize}
\item Given that the baseline Powheg v2~\cite{bib:Powheg} NLO MC simulation of the inclusive $t\bar{t}$ production in both ATLAS and CMS uses the same factorisation and renormalisation scales ($\sqrt{m_{top}^2 + p_{T,top}^2}$), having fixed-order calculations with the same scale for all variables would be helpful; the same choice of PDFs and top mass would help the Collaborations apply the reweighting in a consistent manner (e.g. ATLAS and CMS use a top mass of 172.5 GeV in Powheg v2, while the fixed-order calculations use the value of 173.3 GeV as an input parameter);
\item Given that dedicated fixed-order calculations for the $t\bar{t}$-system mass distribution are available~\cite{bib:Czakon}, a two-dimensional reweighting (i.e. based on top $\pt$ and $t\bar{t}$-system mass) could account for correlations among these variables; an interative, recursive reweighting to these two distributions could be tested as well, giving a MC prediction which matches both top $\pt$ and $t\bar{t}$-system mass fixed-order predictions.

\end{itemize}

\section{Data-driven estimations of the $\ttbar$ background}
The MC simulation-based approach at NLO accuracy in QCD for the prediction of the inclusive $\ttbar$ background is not expected to model well the very high jet and $b$-jet multiplicity regions. Given the current lack of multi-leg calculations, the MC simulation-based approach relies on the description of such large multiplicities through the PS formalism with consequently large uncertainties. Therefore, several data-driven methods have been developed and used to estimate the $\ttbar$ background in regions with large jet and $b$-jet multiplicities and/or high top quark $\pt$. Recent examples of novel estimations of the $\ttbar$ background are presented in Table~\ref{tab:Tab1}. The different data-driven techniques perform well: data agree with the SM expectation within the uncertainties, validating the overall data-driven procedures and the assumptions made, and allowing to search for new physics in extreme phase spaces probed in each of the searches.

\begin{table}[t!]
\caption{A non-comprehensive list of recent ATLAS and CMS searches and measurements applying a data-driven estimation of the $\ttbar$ background.
}
\begin{center}
\begin{tabular*}{\textwidth}{m{11.5em} m{13em} m{12em} } 
  \hline
  \hline
\multicolumn{1}{c}{Analysis / search }
 & \multicolumn{1}{c}{Data-driven technique} &  \multicolumn{1}{c}{Uncertainty}   \\ 
\hline
$Z^{\prime} \rightarrow t\bar{t}$ \newline 0 leptons final state \newline ATLAS Collaboration~\cite{bib:ttRes_ATLAS} 
& Background spectrum derived from data by fitting a smoothly falling function to the mass spectrum of the $\ttbar$ system distribution
& Choice of functional form and the fit range; spurious studies by performing signal-plus-background (S+B) fits constructed under a B-only hypothesis  
 \\ \hline
$X \rightarrow HH \rightarrow b\bar{b}W^{+}W^{-}$ \newline
1 lepton final state \newline CMS Collaboration~\cite{bib:HH_CMS} 
& 
$\ttbar$ background modelled with two-dimensional templates as conditional probabilities using kernel estimation; yields extracted from data using a maximum likelihood fit 
& Morphing the background templates, derived by repeating their construction for different assumptions on jet $\pt$ spectrum and resonance mass \\ \hline
SM and BSM $t\bar{t} t\bar{t}$ \newline 1 and 2 OS leptons \newline ATLAS Collaboration~\cite{bib:my4tops_ATLAS} 
&
Assumption that the probability of $b$-tagging a jet (measured as a function of jet and event properties) in $\ttbar$+jets event is essentially independent of the number of additional jets
&
A full set of systematic uncertainties is derived for the estimate by repeating the procedure on MC simulated events with systematic variations applied
\\ \hline
SUSY $t\bar{t} t\bar{t}$ \newline 1 lepton final state \newline CMS Collaboration~\cite{bib:SUSY4tops_CMS} 
& 
Modified version of an ``ABCD'' method using two uncorrelated variables; any residual correlation taken into account by correction factors
&
Tests performed using data control samples in regions that are kinematically similar but have only a very small potential contribution from signal events.
\\ 
  \hline
  \hline
\end{tabular*}
\label{tab:Tab1}
\end{center}
\end{table}

\subsection{Summary}
Since 2012, the mismodelling of the top quark $\pt$ distribution in data is still one of the main features observed in data. Both ATLAS and CMS analyses correct the top quark $\pt$ in MC by using the reweighting to fixed-order predictions or to differential measurements, and refined approaches to the reweightings are currently possible. The discrepancy seems to be originated due to missing QCD and EW corrections in the MC generators, and tuning efforts may improve the agreement with data.

\medskip
\noindent
In order to search for new physics in extreme phase spaces without dealing with the so-called ``top $\pt$ reweighting'', several data-driven estimation techniques were developed within both ATLAS and CMS Collaborations. These efforts tackle the modelling of the $\ttbar$ background in a new way, and their development should be continued and widely tested.

\end{document}